\documentclass[twocolumn,english,aps,prb,twocolum,superscriptaddress,natbib,bibnotes,amsmath,amssymb,floatfix,groupedaddress,footinbib]{revtex4-2}
\pdfoutput=1
\usepackage[colorlinks=true,citecolor=blue,linkcolor=magenta]{hyperref}

\usepackage[markup=nocolor, authormarkupposition=left]{changes} 

\usepackage{soul}
\usepackage[utf8]{inputenc}
\usepackage[english]{babel}
\usepackage{amsmath,amsfonts,amssymb}
\usepackage[T1]{fontenc}
\usepackage{url}

\usepackage{amsmath}
\usepackage{siunitx}
\usepackage[version=4]{mhchem}
\usepackage{amsfonts}
\usepackage{amssymb}

\usepackage{epstopdf}
\usepackage{graphicx}
\graphicspath{{./Figures/}}

\usepackage{changes}

\begin{document}

\title{Empowering high-dimensional optical fiber communications with integrated photonic processors}

\author{Kaihang Lu$^{1,\ast}$, 
        Zengqi Chen$^{1,\ast}$, 
        Hao Chen$^{1,\ast}$, 
        Wu Zhou$^1$, 
        Zunyue Zhang$^{1}$, 
        Hon Ki Tsang$^{2,\dagger}$, 
        and Yeyu Tong$^{1,\ddagger}$}
\affiliation{
$^1$Microelectronic Thrust, Function Hub, The Hong Kong University of Science and Technology (Guangzhou), Guangdong, PR China\\
$^2$Department of Electronic Engineering, The Chinese University of Hong Kong, Shatin, New Territories, Hong Kong, PR China\\
$^3$School of Precision Instrument and Opto-Electronics Engineering, Tianjin University, Tianjin, PR China\\
}

\maketitle

\noindent\textbf{\noindent
Mode division multiplexing (MDM) in optical fibers enables multichannel capabilities for various applications, including data transmission, quantum networks, imaging, and sensing. However, MDM optical fiber systems, usually necessities bulk-optics approaches for launching different orthogonal fiber modes into the multimode optical fiber, and multiple-input multiple-output digital electronic signal processing at the receiver side to undo the arbitrary mode scrambling in a circular-core optical fiber. Here we show that a high-dimensional optical fiber communication system can be entirely implemented by a reconfigurable integrated photonic processor, featuring kernels of multichannel mode multiplexing transmitter and all-optical descrambling receiver. High-speed and inter-chip communications involving six spatial- and polarization modes have been experimentally demonstrated with high efficiency and high-quality eye diagrams, despite the presence of random mode scrambling and polarization rotation in a circular-core few-mode fiber. The proposed photonic integration approach holds promising prospects for future space-division multiplexing applications. 
}

\section*{Introduction} 

\noindent{The} spatial dimension of multi-mode fibers (MMF) is an unexploited resource for enhancing its information transmission capacity \cite{puttnam_space-division_2021,richardson_space-division_2013,willner_perspectives_2021,van_den_hout_2736_2023}. Space-division multiplexing (SDM), whereby multiple data signals are multiplexed into different spatial channels, has attracted much research interest including the use of multiple cores sharing a common cladding or multiple modes with an enlarged core size and different mode field patterns in an SDM optical fiber. SDM fiber can thus be classified into multi-core fiber (MCF), few-mode fiber (FMF), and MMF \cite{van_uden_ultra-high-density_2014}. Nevertheless, utilizing the spatial dimension of optical fibers can be very challenging, particularly when it involves the higher-order modes in an FMF or MMF for mode-division multiplexing (MDM) systems. Two major challenges are associated with MDM optical fiber systems, including the lack of cost-effective and scalable mode (de)multiplexers that can generate or decouple multiple orthogonal fiber modes, and the substantial energy consumption and large time latency incurred in descrambling high-speed optical signals using digital signal processing \cite{winzer_making_2014}.

Mode (de)multiplexers for optical fibers in several pioneering works have been reported using optical phase plates \cite{ryf_mode-division_2012,shibahara_dmd-unmanaged_2019}, spatial light modulators (SLMs) \cite{gregg_enhanced_2019}, or multi-plane light conversion (MPLC) \cite{rademacher_peta-bit-per-second_2021,labroille_efficient_2014,fontaine_laguerre-gaussian_2019,tang_reconfigurable_2018}. While significant progress has been made in fiber-based photonic lanterns or laser-inscribed waveguides \cite{van_uden_ultra-high-density_2014, fontaine_photonic_2022,gross_ultrafast-laser-inscribed_2015}, a compact and cost-effective approach is still desired for cost- and footprint-sensitive short-reach optical communications inside data centers. Furthermore, when light passes through a circular-core MMF even without any disturbances, various speckle patterns can be formed due to the different dephasing conditions between the fiber eigenmodes \cite{kreysing_dynamic_2014,watanabe_coherent_2020,yang_multi-dimensional_2022,zhang_ultra-high_nodate}. One-by-one mapping of the LP modes between the transmitter side and the receiver side therefore cannot be guaranteed. Intricate speckle patterns at the fiber end can cause significant scrambling of the information encoded on different fiber modes. Rectangular-core FMFs have thus been proposed to break rotational symmetry and prevent spatial degeneracy \cite{yang_multi-dimensional_2022,zhang_ultra-high_nodate}. In addition, inter-mode coupling may also arise from factors such as fiber non-uniformity or sharp bending. Essentially, the information is not lost, but separating the arbitrarily mixed signals in a circular-core FMF presents a significant challenge. Prior research has shown that this problem can be well solved using digital electronic multiple-input multiple-output (MIMO) processing \cite{ryf_mode-division_2012,winzer_mimo_2011}. However, such approaches originally developed for wireless communications require high-speed digital circuits with high power consumption and large time latency when handling a data rate of over 100Gbaud, which severely hinders its potential use.

\begin{figure*}[!htp]
  \centering{
  \includegraphics[width = 0.97\linewidth]{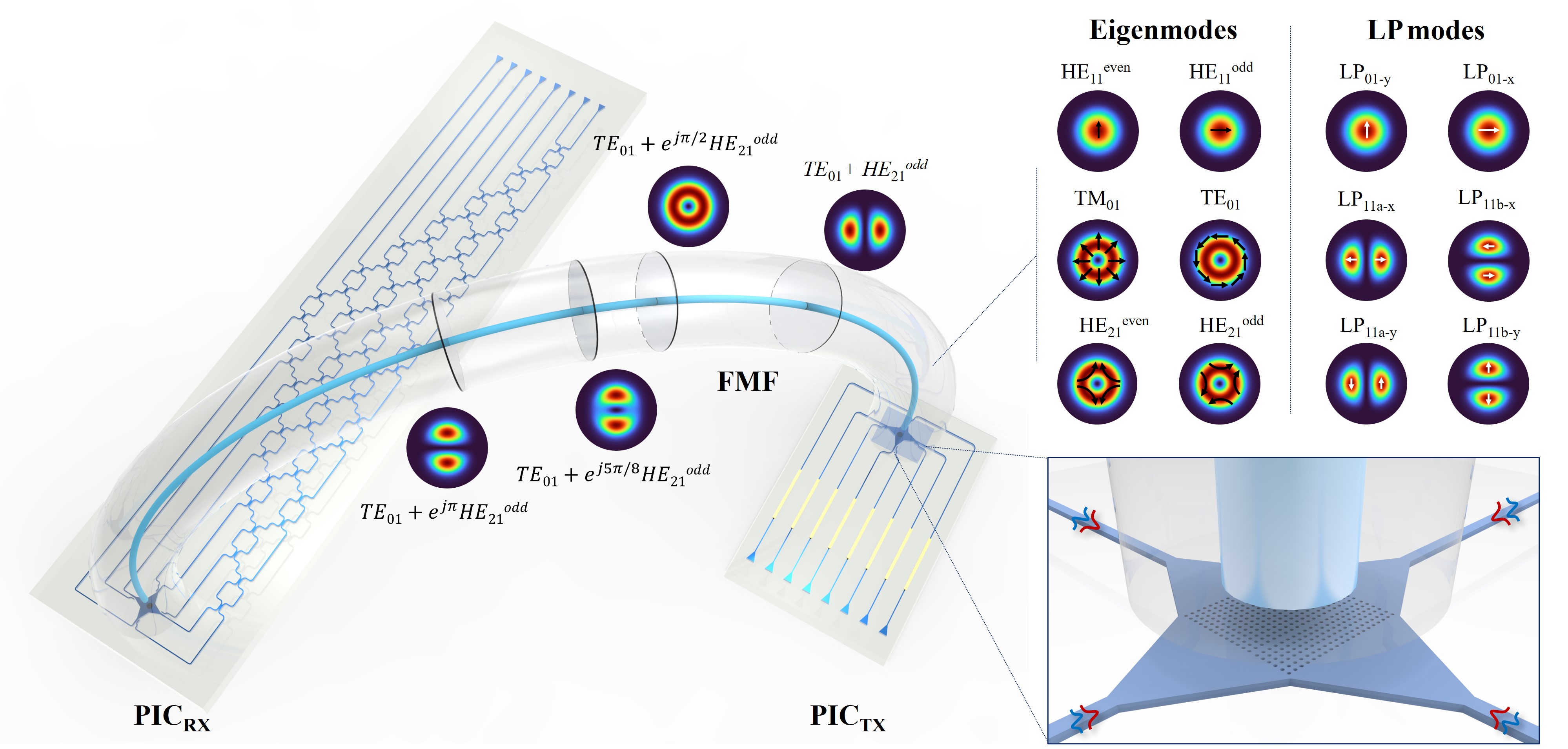}
  } 
    \caption{\noindent\textbf{Schematic of the high-dimensional optical fiber communication system with reconfigurable integrated photonic processor.} Top inset: mode field profile of the eigenmodes and linear polarized (LP) modes in a two-mode few-mode fiber (FMF). Bottom inset: ariel view of the integrated multimode grating coupler as optical I/O (not in scale). PIC\textsubscript{TX}: photonic integrated circuits at the transmitter side; PIC\textsubscript{RX}: photonic integrated circuits at the receiver side. Continuous mixing of optical signals occurs during propagation in a circular-core FMF due to rotation symmetry and spatial degeneracy.
  }
 \label{Figure1}
\end{figure*} 

Integrated photonic processors present a promising alternative technology for manipulating the high-order fiber modes, especially on the silicon photonics platform, which offers low-cost, high-volume manufacturing with CMOS compatibility \cite{bogaerts_programmable_2020,miller_device_2009,miller_perfect_2015}. The high refractive index contrast of the silicon photonics platform enables ultra-compact confinement of the optical field for high-density and multichannel optical input/output (I/O) \cite{watanabe_coherent_2020,tong_efficient1_2020,jimenez_gordillo_fiber-chip_2023,
zhou_high_2022,tong_efficient_2018}. Meanwhile, integrated optical interferometers to implement arbitrary matrix transformations have been demonstrated for use in optical neural networks \cite{shen_deep_2017,zhang_optical_2021,shastri_photonics_2021}, reconfigurable signal processors \cite{huang_silicon_2021,zhang_broadband_2023,zhou_all--one_2019,tang_ten-port_2021}, free-space and on-chip beam separation \cite{annoni_unscrambling_2017,
milanizadeh_separating_2022,wu_chip--chip_2023}, and quantum networks \cite{zheng_multichip_2023,carolan_universal_2015}. By integrating multimode optical I/O and optical matrix processing on the same chip, photonic processors have the potential to offer an enabling technology for MDM optical fiber systems.

In this work, we developed a reconfigurable integrated photonic processor that can selectively launch and separate orthogonal fiber modes. High-dimensional chip-to-chip optical fiber communications can be directly realized by an FMF, involving the full set of six LP modes including LP\textsubscript{01-x}, LP\textsubscript{01-y}, LP\textsubscript{11a-x}, LP\textsubscript{11a-y}, LP\textsubscript{11b-x}, and LP\textsubscript{11b-y}. Selective mode excitation in the optical fiber is performed at the transmitter side by an efficient and multimode optical I/O. To solve the mode scrambling and polarization rotation after fiber transmission at the receiver side, a reconfigurable Mach Zehnder interferometer (MZI) based optical unitary mesh is utilized to apply optical matrix transformations and act as an all-optical MIMO descrambler. For the first time, a six-channel high-dimensional fiber optical communication relying on the integrated silicon photonic processor is experimentally demonstrated.

\section*{Integrated photonic processor} 
Figure \ref{Figure1} illustrates the high-dimensional optical fiber communication system enabled by the silicon photonic integrated circuits (PICs). The higher-order modes in an FMF can be described by eigenmodes with a rigorous vectorial treatment of the wave equation in cylindrical coordinates. LP mode is a description often used for the linearly polarized superposition of fiber eigenmodes. At the transmitter side, the chip-to-fiber coupling is realized by an efficient and multimode grating coupler as depicted by the bottom inset. By controlling the relative phase delay between the two counterpropagating quasi-transverse-electric (TE) modes on chip, high-order fiber modes in the two orthogonal polarizations can be generated as elaborated in Supplementary Material S1. In our demonstration, all the six spatial- and polarization channels in a two-mode FMF can be selectively launched, including LP\textsubscript{01-x}, LP\textsubscript{01-y}, LP\textsubscript{11a-x}, LP\textsubscript{11a-y}, LP\textsubscript{11b-x}, and LP\textsubscript{11b-y}. It is worth noting that although selective decoupling is desired at the receiver side, this can only happen when the fiber LP modes are of high modal purity and polarizations are accurately aligned. In practice, LP mode scrambling and polarization rotation are inevitable in a circular-core FMF, which results in an unpredictable field pattern arriving at the coupling end between the fiber and the photonic chip\cite{kreysing_dynamic_2014,watanabe_coherent_2020,yang_multi-dimensional_2022}. For example, when the LP\textsubscript{11a} mode is launched into the FMF from the transmitter side, the resulting output pattern is typically a linear combination of all LP\textsubscript{11} spatial- and polarization modes, which undergo continuous mixing along propagation in the optical fiber, as illustrated in Figure 1. This is due to the fact that LP modes are essentially formed by linear combinations of the fiber eigenmodes with varying interference conditions. Nevertheless, the proposed multimode grating coupler can efficiently support all the eigenmodes in a two-mode FMF, which allows non-selective decoupling of the multimode optical signals into the eight single-mode channels on chip. Efficient chip-to-fiber and fiber-to-chip coupling can be obtained with a small mode-dependent loss as shown by the simulation results in Figure S2. To undo the arbitrary inter-modal signal mixing, reconfigurable optical mesh with eight output ports can be trained to apply the inverse transformation of the channel matrix and retrieve the six orthogonal channels in a two-mode FMF. 

 \begin{figure*}[!htp]
  \centering{
  \includegraphics[width = 0.97\linewidth]{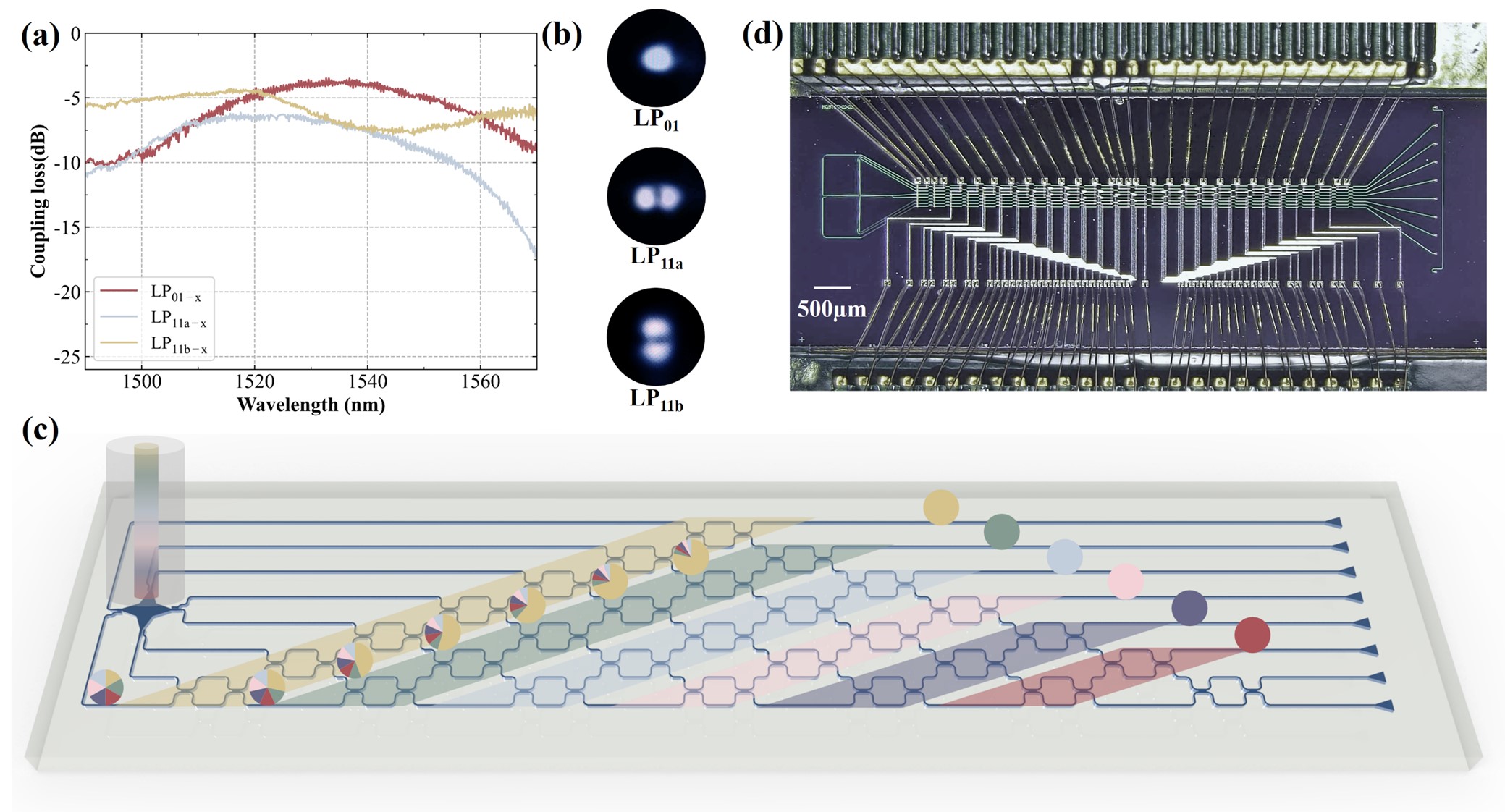}
  } \caption{ (a) Experimental chip-to-fiber coupling loss spectra of the multimode grating coupler for various LP modes. (b) Field profile of the FMF captured by an infrared camera with a 10× microscope objective when different fiber mode is selectively launched by the PIC\textsubscript{TX}. Schematic of the optical mesh based on Mach-Zehnder interferometers (MZIs) at the receiver side for mode unscrambling. (d) Microscopic image of the wired-bonded photonic integrated circuits used at the received side.}
  \label{Figure2}
\end{figure*} 

 In this demonstration, a unitary optical mesh \textit{U(8)} is applied leveraging the low channel-dependent loss of the multimode optical I/O. More intricate optical meshes would be necessary to enhance the system performance with severe loss difference or differential mode group delay\cite{miller_establishing_2013,zhang_-chip_2023}.  A comprehensive explanation of the unitary matrix operations of the photonic processor is included in the Supplementary Material (refer to Section S2).

 The experimental coupling loss spectra of LP\textsubscript{01}, LP\textsubscript{11a}, and LP\textsubscript{11b} in x-polarization for a two-mode FMF are measured and presented in Figure \ref{Figure2}a, showing a peak experimental efficiency of -3.5 dB, -6.1 dB, and -4.3 dB at 1532nm, 1517nm, and 1515nm respectively.  As the grating utilizes a symmetric structure for the orthogonal polarizations, similar coupling efficiencies for the y-polarized LP modes are guaranteed. To validate the selective launching of the LP\textsubscript{01}, LP\textsubscript{11a}, and LP\textsubscript{11b} modes when using the integrated photonic processor, diffracted mode field profiles of FMF are captured with a 10 $\times$ microscope objective and an infrared camera, as presented in Figure \ref{Figure2}b. The FMF is twisted and bent by using the fiber loops until the captured image resembles the profile of a pure LP\textsubscript{11} mode. The schematic of the photonic processor at the receiver side is shown in Figure \ref{Figure2}c. It comprises a multimode grating coupler, tapered asymmetric directional couplers (ADCs), MZI-based linear unitary matrix, and eight output single-mode grating couplers. An 8$\times$8 triangular optical mesh consisting of 28 tunable MZIs is employed in our demonstration. Figure \ref{Figure2}d shows the microscopic image of wire-bonded photonic processor at the receiver side. The total footprint of the photonic integrated circuits is 8.5 mm $\times$ 1.8 mm. 

\section*{Mode descrambling training} 
 \begin{figure*}[htp]
  \centering{
  \includegraphics[width = 0.97\linewidth]{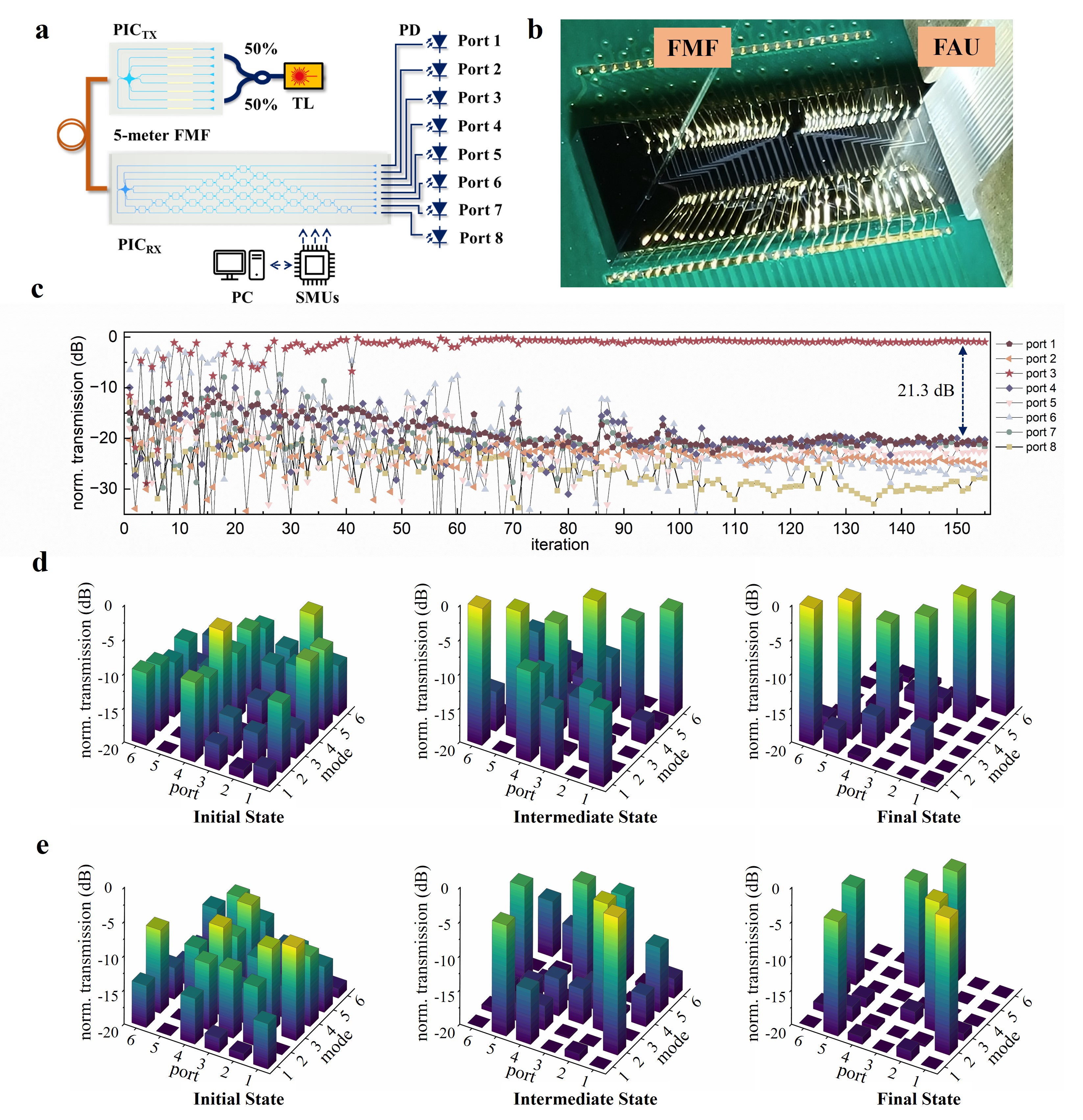}
  } \caption{(a) Experimental setup used for inter-chip MDM fiber communication system. TL: tunable laser, BS: beam splitter, FMF: few-mode fiber, PD: photodiode. (b) Photograph of the wired-bonded photonic chip under test at the receiver side with FAU (fiber array unit) and FMF. (c) Evolution of the normalized transmission for eight output ports with the training process. (d) Bar chart of the initial random state, intermediate state, and final state for 6 space and polarization channels during the training process. (e) Bar chart of another routing configuration after training by the PSO algorithm. The mode 1-6 refer to LP\textsubscript{01-x}, LP\textsubscript{01-y}, LP\textsubscript{11a-x}, LP\textsubscript{11a-y}, LP\textsubscript{11b-x}, LP\textsubscript{11b-y}.}
  \label{Figure4}
\end{figure*} 

 \begin{figure*}[htp]
  \centering{
  \includegraphics[width = 0.97\linewidth]{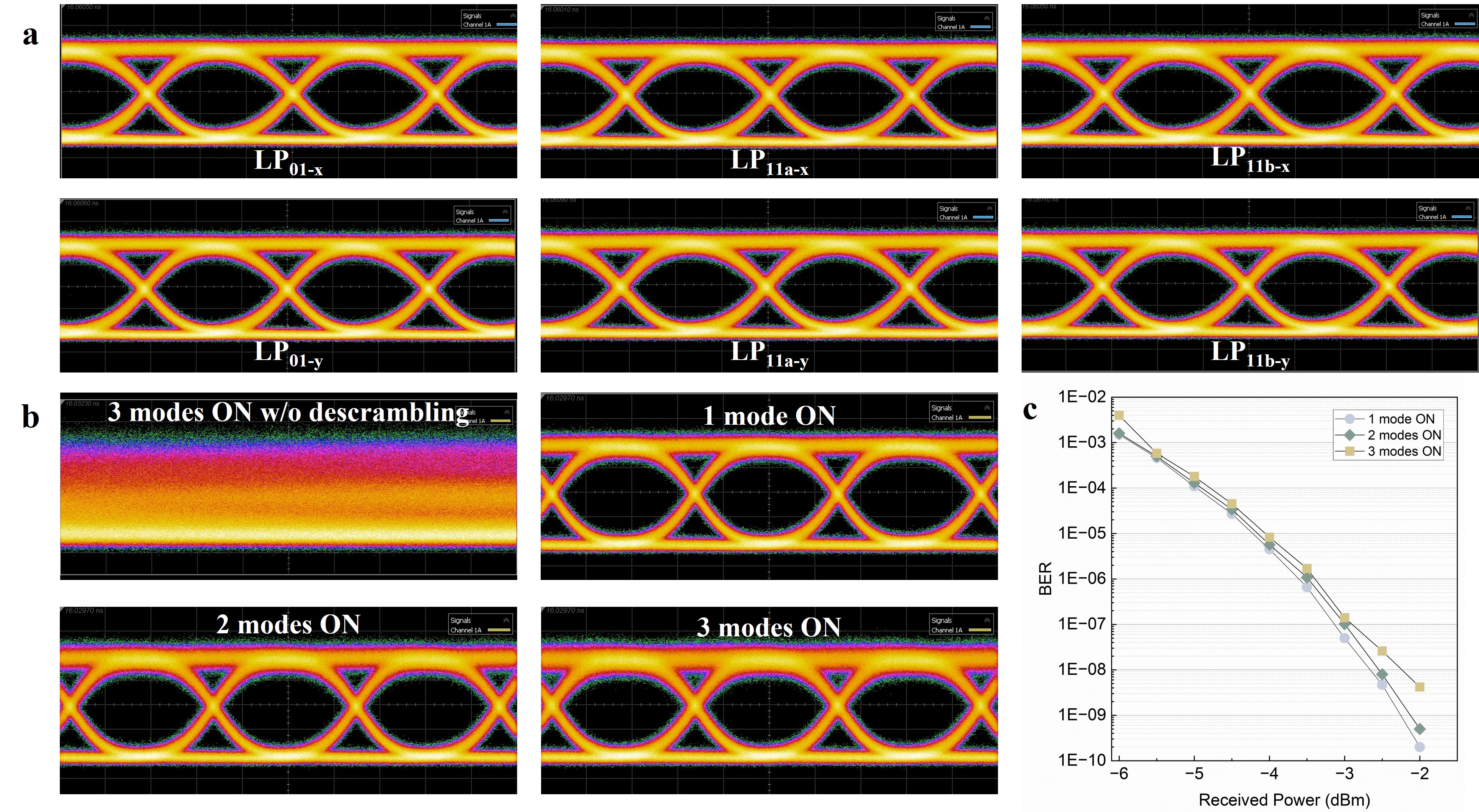}
  } \caption{(a) 32Gbps eye diagrams of the six space and polarization channels retrieved by the integrated photonic processor. (b) Eye diagrams and (c) BERs retrieved and with a minor penalty when additional concurrent fiber modes are turned on. The eye diagram is completely closed without mode reconstruction due to the coherent beating of the three concurrent modes.}
  \label{Figure5}
\end{figure*} 
With various fiber modes selectively launched, the FMF is directly bridged with the photonic chip at the receiver side as indicated by Figure \ref{Figure4}a. Because of the mode evolution and polarization rotation, the received speckle pattern at the fiber-to-chip end is uncertain, which results in a random received optical power distribution entering the optical mesh. A fiber array unit (FAU) and multichannel power meter are used to monitor the optical power from the eight output ports. Figure \ref{Figure4}a shows the experimental setup for chip-to-chip MDM communications using a 5-meter FMF and the integrated photonic processor. The photograph of the photonic chip mounted on a printed circuit board at the received side under test is presented in Figure \ref{Figure4}b. In the experiments, we have employed multichannel source measurement units (SMUs) and the particle-swarm optimization (PSO) algorithm to optimize the drive voltage of phase shifters within the optical mesh \textit{U(8)}. The figure-of-merits has been defined as the minimum crosstalk suppression between the target output port and the other output ports. Once an acceptable solution is identified or the maximum number of iterations is reached, the optimization algorithm terminates and locks the unscrambling status by fine-tuning the driving voltage of each phase shifter.

Figure \ref{Figure4}c presents the evolution of normalized optical power for eight output ports during the training process. It can be clearly observed that after a training epoch of around 120, selective decoupling can be effectively configured with a maximum crosstalk level suppression of $\ge$ 21.3 dB. Through the same experimental setup, six channels including the two orthogonal polarizations of the LP\textsubscript{01}, LP\textsubscript{11a}, and LP\textsubscript{11b} modes are selectively launched and decoupled with transmission matrices bar chart summarized in Figure \ref{Figure4}d. The initial power randomization for all fiber modes can be configured to different output ports, with experimental power isolation ratios all above 15.2 dB. Different routing schemes of fiber modes are also implemented to validate the reconfigurability of our photonic processor, showing a similar performance as depicted in Figure \ref{Figure4}e. A comprehensive optical loss breakdown of the entire optical system is in Supplementary Material S3. 

\section*{Inter-chip multimode communication}

To evaluate the high-speed communication performance of the reconfigurable photonic processor, 32Gbps non-return-to-zero (NRZ) signals are generated at a wavelength of 1530 nm. The pseudorandom binary sequence (PRBS) is obtained with a bit pattern generator (BPG) and a LiNbO\textsubscript{3} Mach-Zehnder modulator. The experimental setup is shown in Figure S4a, various orthogonal fiber modes including LP\textsubscript{01-x}, LP\textsubscript{01-y}, LP\textsubscript{11a-x}, LP\textsubscript{11a-y}, LP\textsubscript{11b-x}, and LP\textsubscript{11b-y} are selectively launched in the FMF at the transmitter side. After the training process at the receiver side, the photonic processor can route the modulated optical signals to any desired output port and undo the signal mixing process. Figure \ref{Figure5}a presents clear and open eye diagrams for each of the fiber modes which primarily benefited from the low-loss multichannel optical I/O. The experimental results reveal that a high-dimensional optical fiber communication system can be realized by our reconfigurable integrated photonic processor.

To further evaluate the all-optical descrambling performance at the receiver side with additional concurrent channels injected, three spatially decoupled data channels are launched simultaneously from the transmitter side. It is worth mentioning that a mode-selective fiber photonic lantern \cite{leon-saval_mode-selective_2014} is used here, because the input grating couplers of our transmitter are not optimally positioned for an FAU in multichannel operation (as shown in Figure S4b), which can be addressed in the future design. The experimental setup with three concurrent channels injected is illustrated in Figure S4c. Due to the coherent interference between the three spectrally overlapped channels, the corresponding eye diagram is completely closed, as shown in Figure \ref{Figure5}b. In contrast, when the optical mesh is carefully configured and trained, a clear and open eye diagram can be effectively reconfigured. As more concurrent data channels are activated, the eye diagram experiences very slight degradation due to a crosstalk suppression level of >21 dB. Bit error rate measurements are also performed as presented in Figure \ref{Figure5}c, which shows a power penalty of about 0.5 dB at a BER threshold of 10\textsuperscript{-8} when all three concurrent modes are switched on.

\section*{Discussion}

In summary, we have proposed and demonstrated a high-dimensional optical fiber communication system which is enabled by a reconfigurable integrated photonic processor. Selective mode launching and all-optical mode descrambling were realized by the integrated multimode optical I/O and MZI-based optical unitary mesh. The photonic processor is capable of unscrambling the random mode and polarization mixing that occurs during transmission through a circular-core FMF. Six spatial- and polarization channels, including the full set modes in the LP\textsubscript{01} and LP\textsubscript{11} mode groups can be utilized for the MDM fiber communications with high transmission efficiency. High-quality eye diagrams can be retrieved even when additional concurrent modes are activated, showing a very small power penalty on the measured BERs.

To further reduce the residual crosstalk unscrambled by the integrated optical mesh, one approach is to optimize the center wavelength of the multimode optical I/O to decrease channel-dependent loss. Another universal solution is to implement more complex optical mesh architectures to execute non-unitary linear matrix transformations\cite{miller_establishing_2013,
zhang_-chip_2023}. This approach can descramble optical modes that have experienced substantial mode-dependent loss and differential mode group delay, at the expense of increased circuit complexity. Additionally, to prevent communication interruptions, progressive self-configuration with feedback can also be implemented for simple and precise control of the integrated optical mesh\cite{annoni_unscrambling_2017,
milanizadeh_separating_2022}, and to handle the arbitrary mode evolution and polarization rotation during the long-distance optical fiber transmission.

The number of involved spatial channels can also be increased by optimizing the multimode optical I$/$O and the optical mesh. For instance, LP\textsubscript{21b} mode can also be launched by feeding two counterpropagating TE1 modes with a relative phase shift of $\pi$, using the same multimode grating coupler at the transmitter side in this work. To undo the signal mixing, the receiver side would require a non-unitary optical mesh as not all of the degenerate modes in the LP\textsubscript{21} group can be efficiently coupled back to the photonic chip. This would also necessitate the scaling of the dimension size of the integrated optical mesh and optical I$/$O. Silicon photonic integrated processors are expected to be an enabling technology for future MDM applications in a variety of different fields, such as communications, imaging, sensing, and quantum networks.

\vspace{0.5cm}

\noindent\textbf{Methods}
\medskip
\begin{footnotesize}

\noindent\textbf{Multimode optical I/O design.} The multimode grating coupler is designed for operation around 1550 nm wavelength range with a perfect vertical coupling configuration. 70-nm shallow etched holes are utilized as the low-index region for diffraction with a symmetrical pattern for the orthogonal polarizations. To reduce the coupling loss, chirped grating periods and hole diameter are optimized by genetic algorithm with effective medium theory and 2 dimensioanl (2D) finite-difference time-domain (FDTD) simulations. 3D FDTD simulations are performed to validate the coupling performances of all the high-order fiber modes. Four asymmetrical directional couplers (ADCs) are used to (de)multiplex the TE\textsubscript{0} and TE\textsubscript{1} modes on chip. The relative phase shift is adjusted by a heater-based waveguide phase shifter with titanium-tungsten alloy (TiW) on the top. The two-mode FMF used in our experiment is fabricated by OFS. 

\vspace{0.1cm}

\noindent\textbf{Photonic chip fabrication.} The photonic chip is fabricated on a silicon-on-insulator (SOI) wafer with a 220 nm thick top silicon layer. The buried-oxide layer is 2 µm thick. Electro-beam lithography is used to define the device patterns, followed by dry reactive-ion etching process with a etch depth of 70 nm and a full etch. To protect the photonic circuits, a top cladding of silicon dioxide (SiO\textsubscript{2}) with a thickness of 1.2 µm is used. Metallization is done using high-resistance titanium-tungsten alloy (TiW) for local heat generation and aluminum for electrical signal routing. A 300-nm thick SiO\textsubscript{2} passivation layer is used and selectively etched later over the aluminum pads for probing.

\vspace{0.1cm}

\noindent\textbf{MDM chip-to-chip data transmission.} For high-speed inter-chip optical communications, the tunable laser (Santec TSL-570) is set at 1530 nm with an output power of 13dBm injected into a LiNbO\textsubscript{3} Mach-Zehnder modulator (AFR AM40). The optical modulator is driven by a BPG (Keysight 8045A) with an RF amplifier (SHF S807C). The modulator optical signals are spatially decoupled by SMFs with a transmission distance of 2 km and 5 km and sent to the mode-selective fiber photonic lantern with a two-mode FMF. The integrated photonic processor is controlled by a multichannel source measurement unit (Nicslab XDAC-120U-R4G8) and personal computer for the training process. An 8-channel optical power meter (Santec MPM-210H and MPM-215) is employed to read the optical powers from the fiber array. For eye diagram characterization at the receiver side, the optical signal is boosted by an erbium-doped fiber amplifier (EDFA, Amonics AEDFA-PA-35-B-FA) and sent to a 50-GHz PIN photodiode (Coherent XPDV2320R). The eye-diagram and bit error rate are obtained from a sampling oscilloscope (Keysight N1000A) and BERT (Keysight 8040A).

\vspace{0.1cm}

\noindent \textbf{Acknowledgements}:
This work was supported by Guangzhou - HKUST(GZ) Joint Funding Program (No. 2023A03J0159), Start-up fund from the Hong Kong University of Science and Technology (Guangzhou), and Hong Kong Innovation and Technology Fund project ITS/226/21FP. The authors acknowledge the Novel IC Exploration (NICE) Facility of HKUST(GZ) for device measurement and Applied Nanotools Inc. for device fabrication.

\vspace{0.1cm}

\noindent \textbf{Conflicts}:
The authors declare no conflicts of interest.
\end{footnotesize}

\renewcommand{\bibpreamble}{
$^\ast$These authors contributed equally to this work.\\
$^\dagger${Corresponding author: \textcolor{magenta}{hktsang@ee.cuhk.edu.hk}}\\
$^\ddag${Corresponding author: \textcolor{magenta}{yeyutong@hkust-gz.edu.cn}}
}
\bibliographystyle{naturemag}
\bibliography{0_main}

\end{document}


\title{Supplementary information for:\\ Empowering high-dimensional optical fiber communications with integrated photonic processers}

\author{Kaihang Lu$^{1,\ast}$, 
        Zengqi Chen$^{1,\ast}$, 
        Hao Chen$^{1,\ast}$, 
        Wu Zhou$^1$, 
        Zunyue Zhang$^{1}$, 
        Hon Ki Tsang$^{2,\dagger}$, 
        and Yeyu Tong$^{1,\ddagger}$}
\affiliation{
$^1$Microelectronic Thrust, Function Hub, The Hong Kong University of Science and Technology (Guangzhou), Guangdong, PR China\\
$^2$Department of Electronic Engineering, The Chinese University of Hong Kong, Shatin, New Territories, Hong Kong, PR China\\
$^3$School of Precision Instrument and Opto-Electronics Engineering, Tianjin University, Tianjin, PR China\\
}
\maketitle

\noindent{\textbf{\large{Contents}}}

\noindent{\textbf{Supplementary Notes:}}\\
\noindent{1. Multimode Optical I/O for Few Mode Fiber}\\
2. Mathematics and Matix Operation of the Integrated Photonic Processor\\
3. Optical Loss Analysis of Integrated Photonic Processor\\
4. Chip-to-Chip Mode Division Multiplexing Optical Communication

\noindent{\textbf{Supplementary Figures:}}\\
\noindent{Figure S1. Schematic of integrated multimode optical I/O}\\
Figure S2. Chip-to-fiber and fiber-to-chip coupling efficiency spectra in simulaton\\
Figure S3. Unitary optical mesh\\
Figure S4. Experimental setup for high speed communications\\

\section*{\textbf{Supplementary Note 1. MULTIMODE OPTICAL I/O FOR FEW MODE FIBER}}

\noindent{The} multimode optical I/O consists of a two-dimensional (2D) multimode grating coupler, four linearly tapered mode size converters, and four tapered asymmetrical directional couplers (ADCs)\cite{ding_-chip_2013,dai_silicon_2013} as illustrated by Figure S1a. Selective fiber mode launching at the transmitter can be realized via controlling the relative phase difference between the two counterpropagating quasi-transverse-electric (TE) modes. Since the fundamental (TE\textsubscript{0}) and first-order (TE\textsubscript{1}) modes exhibit similar effective indices in the silicon gratings with a width of 13 µm, they can be efficiently diffracted out of the plane relying on the same grating structure. The linearly polarized (LP) modes in a two-mode few-mode fiber (FMF) include LP\textsubscript{01-x}, LP\textsubscript{01-y}, LP\textsubscript{11a-x}, LP\textsubscript{11a-y}, LP\textsubscript{11b-x}, and LP\textsubscript{11b-y}. The corresponding launching conditions for each optical fiber mode are summarized in Figure S1b. To guarantee the same coupling performance for all the optical modes coming from the four orthogonally placed waveguides, the grating diffraction region is centrosymmetric with chirped grating periods optimized by a genetic optimization algorithm and finite-difference time-domain (FDTD) simulations.

The schematic of the tapered ADC for on-chip TE\textsubscript{0}-TE\textsubscript{1} mode (de)multiplexing is shown by Figure S1c. The integrated waveguide width w\textsubscript{1}, w\textsubscript{2a}, and w\textsubscript{2b} are 0.45 µm, 0.902 µm, and 0.962 µm, respectively. The waveguide gap g is designed as 0.2 µm and coupling length L is 33.6 µm. To match the mode field diameter between the diffracted beam and fiber modes in the two-mode FMF fabricated by \textit{OFS}, the width of silicon waveguide is linearly tapered from 0.962 µm to 13 µm by an adiabatic linear taper with a length of L\textsubscript{taper}=350 µm as shown in Figure S1a.

The simulated chip-to-fiber coupling efficiency spectra of the 2D multimode grating coupler is shown in Figure S2a. The x-polarized LP\textsubscript{01}, LP\textsubscript{11a}, and LP\textsubscript{11b} modes manifest a similar coupling efficiency of about –4.6 dB at a wavelength of 1558 nm. The wavelength span is about 21 nm with a mode-dependent loss is less than 3 dB. At the receiver side, because of random mode evolution and polarization rotation in a circular core FMF, the received speckle pattern is uncertain at the fiber end. However, the multimode optical I/O can support efficient coupling for all the eigenmodes in a two-mode FMF. To validate that, fiber-to-chip coupling simulation is performed by launching various eigenmodes in a two-mode FMF and summing the received optical power into TE\textsubscript{0} and TE\textsubscript{1} modes on chip. The coupling loss spectra for HE\textsubscript{11-x}, HE\textsubscript{11-y}, TM\textsubscript{01}, TE\textsubscript{01}, HE\textsubscript{21-even}, and HE\textsubscript{21-odd} as shown in Figure S2b. The two simulation results show good agreement with a minor discrepancy, suggesting that the multimode optical I/O can be utilized at both the transmitter and receiver sides, with small mode-dependent loss.

\section*{\textbf{{Supplementary Note 2. Mathematics and Matrix Operation of the Integrated Photonic Processor}}}

The photonic integrated processor at the receiver side employs an 8$\times$8 unitary optical mesh \textit{U(8)} to undo the arbitrary signal mixing among the complete set of 6 orthogonal channels inside of a two-mode FMF. The optical mesh is composed of a triangular array of 28 tunable 2$\times$2 beam splitters (BSs). As shown in Figure S3, each beam splitter forms simple 2$\times$2 unitary transformations consisting of two heater-based optical phase shifters and two multimode interferometers (MMIs). The corresponding scattering matrix of each BS can be obtained from the product of the subcomponents shown by Equation S1, where $\theta_1$ and $\theta_2$ are the outer and inner optical phase shifts obtained by the tunable heaters. Equation S1 can be simplified to Equation S2 by applying Euler’s equation. Unitary transformation \textit{U(2)} can thus be realized as indicated by Equation S3. An 8$\times$8 unitary transformation matrix can be formed by cascading the tunable BSs in a triangular topology as depicted by Equation S4.
\begin{equation}
S=\left( \begin{matrix}
   \frac{1}{\sqrt{2}} & \frac{j}{\sqrt{2}}  \\
   \frac{j}{\sqrt{2}} & \frac{1}{\sqrt{2}}  \\
\end{matrix} \right)\left( \begin{matrix}
   1 & 0  \\
   0 & {{e}^{j{{\theta }_{2}}}}  \\
\end{matrix} \right)\left( \begin{matrix}
   \frac{1}{\sqrt{2}} & \frac{j}{\sqrt{2}}  \\
   \frac{j}{\sqrt{2}} & \frac{1}{\sqrt{2}}  \\
\end{matrix} \right)\left( \begin{matrix}
   1 & 0  \\
   0 & {{e}^{j{{\theta }_{1}}}}  \\
\end{matrix} \right)\
\end{equation}
\begin{equation}
S=j{{e}^{j\frac{{{\theta }_{2}}}{2}}}\left( \begin{matrix}
   -\sin \frac{{{\theta }_{2}}}{2} & {{e}^{j{{\theta }_{1}}}}\cos \frac{{{\theta }_{2}}}{2}  \\
   \cos \frac{{{\theta }_{2}}}{2} & {{e}^{j{{\theta }_{1}}}}\sin \frac{{{\theta }_{2}}}{2}  \\
\end{matrix} \right)\
\end{equation}
\begin{equation}
S{{S}^{\dagger }}={{S}^{\dagger }}S=E\
\end{equation}
\begin{equation}
U(N)=\prod\limits_{i=2}^{N}{\left[ \prod\limits_{j=1}^{i-1}{{{T}_{ij}}} \right]}\
\end{equation}
where T\textsubscript{ij} is shown in Equation S5, i is from 2 to \textit{N}, \textit{j = i – 1}. Each BS changes a two-dimensional subspace of Equation S4.
\begin{equation}
    {{T}_{ij}}=\left( \begin{matrix}
   1 & 0 & \cdots  & \cdots  & \cdots  & \cdots  & 0  \\
   0 & 1 & \cdots  & \cdots  & \cdots  & \cdots  & 0  \\
   \vdots  & \vdots  & \ddots  & \cdots  & \cdots  & \cdots  & \vdots   \\
   0 & 0 & \cdots  & -\sin \frac{{{\theta }_{2}}}{2} & {{e}^{j{{\theta }_{1}}}}\cos \frac{{{\theta }_{2}}}{2} & \cdots  & 0  \\
   0 & 0 & \cdots  & \cos \frac{{{\theta }_{2}}}{2} & {{e}^{j{{\theta }_{1}}}}\sin \frac{{{\theta }_{2}}}{2} & \cdots  & 0  \\
   \vdots  & \vdots  & \cdots  & \cdots  & \cdots  & \ddots  & \vdots   \\
   0 & 0 & \cdots  & \cdots  & \cdots  & \cdots  & 1  \\
\end{matrix} \right)
\end{equation}

\section*{\textbf{Supplementary Note 3. Optical Loss Analysis of Integrated Photonic Processor}} 

In the experiment, the chip-to-chip high-speed optical communications are performed at a wavelength of 1530 nm to minimize the channel-dependent loss, while within the working wavelength range of the erbium-doped fiber amplifier (EDFA) for optical amplification.

At the transmitter side, each single-mode input grating coupler experiences a transmission loss of approximately 5.5 dB. The multimode grating coupler has a coupling loss of about 4$\sim$7 dB for LP\textsubscript{01} and LP\textsubscript{11} mode groups, respectively. The optimal insertion loss of the transmitter side is approximately 10 dB, with a loss variation of roughly 3 dB. In the future, the insertion loss and loss variation can be improved by optimizing the center wavelength of the input single-mode grating coupler and multimode grating coupler.

At the receiver side, the optical loss can be broken down into 4$\sim$7 dB from the multimode grating coupler, 5.5 dB from the single mode output grating coupler, around 0.35 dB for each multimode interferometer (MMI) (center wavelength is designed at 1550 nm), about 4.5 dB from the waveguide and bends inside the optical mesh. The optimal total insertion loss of the receiver side is approximately 15 dB, with a loss variation within 6 dB due to different routing paths with varying numbers of MMIs and the mode-dependent loss of the multimode grating coupler. In the future, the total loss and loss variation can be mitigated by aligning the center working wavelength of the MMI to the multimode grating coupler.

\section*{\textbf{Supplementary Note 4. Chip-to-Chip Mode Division Multiplexing Optical Communication}}

A chip-to-chip high-dimensional optical communication system is established by the integrated photonic processor. Figure S4a presents the experimental setup for the high-speed eye diagram measurements when different fiber modes are selectively launched from the transmitter side, and descrambled by the integrated optical mesh at the receiver side after the training process. We also evaluate the mode descrambling performance at the receiver side with additional orthogonal modes turned on. Since the input grating coupler array of the integrated transmitter is not compatible with a fiber array unit with channel pitch of 250 µm, a fiber-based mode selective photonic lantern\cite{leon-saval_mode-selective_2014} is utilized as shown in Figure S4b. LP\textsubscript{01}, LP\textsubscript{11a}, and LP\textsubscript{11b} modes are spatially decoupled first by 2-km and 5-km SMFs before launching into the fiber photonic lantern. The photonic processor at the receiver side will be trained to reconstruct high-speed eye diagrams for one of the spatial channels. Once completed, the other two spatial channels will be activated to assess the induced penalty on the eye diagrams and bit error rates. 

\clearpage

\begin{figure*}[!hp]
  \renewcommand{\figurename}{Supplementary Figure}
  \centering{
  \includegraphics[width = 1.0\linewidth]{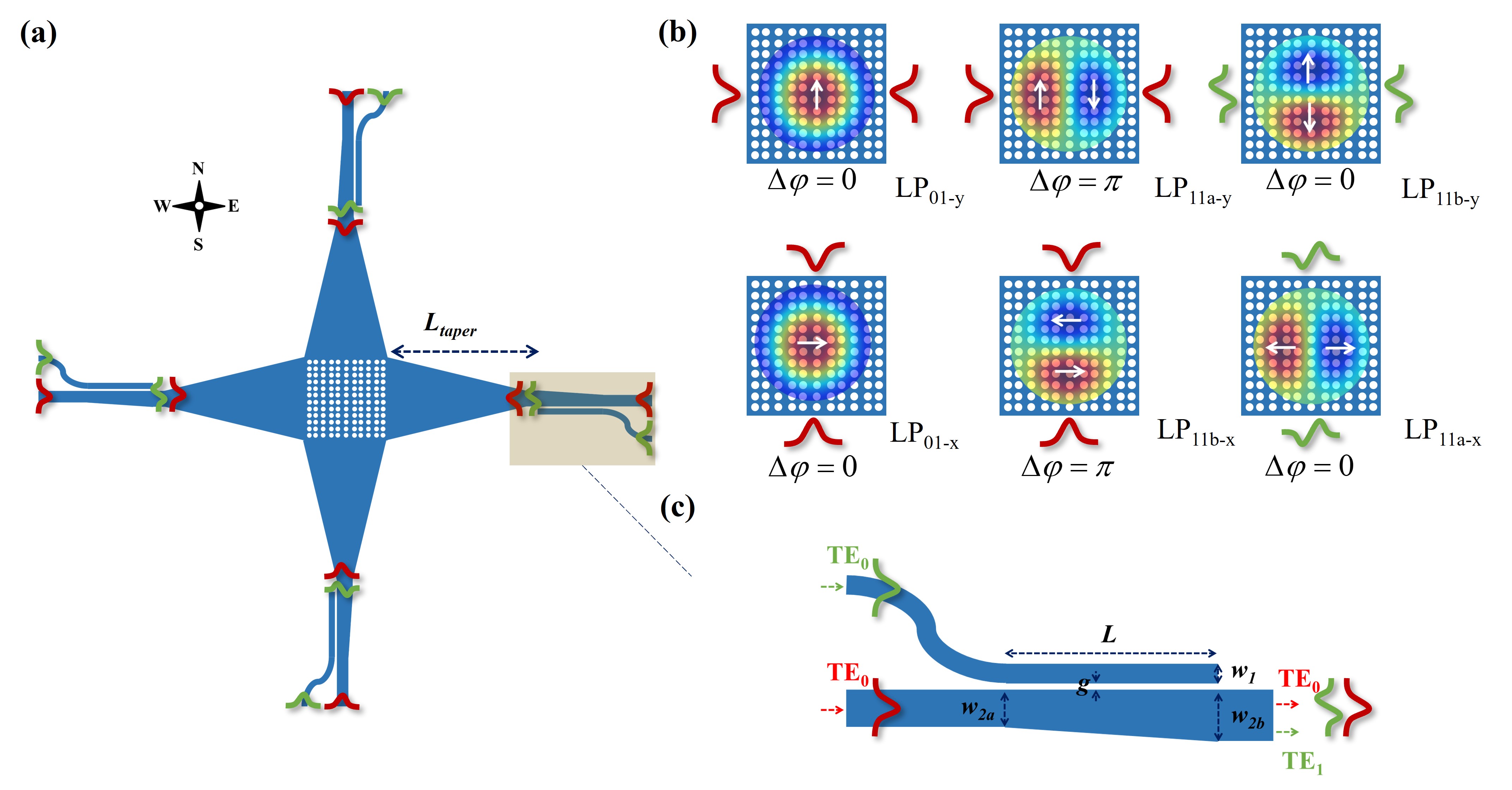}
  } 
  \caption{(a) Schematic of integrated multimode optical I/O, consisting of a two-dimensional waveguide grating, tapered adiabatic directional couplers, and asymmetrical directional couplers. (b) Illustration of fiber mode (de)multiplexing using the multimode grating coupler. (c) Schematic of the asymmetrical directional coupler for TE\textsubscript{0}-TE\textsubscript{1} mode (de)multiplexing.}
 \label{FigureS1}
\end{figure*} 

\begin{figure*}[!hp]
  \renewcommand{\figurename}{Supplementary Figure}
  \centering{
  \includegraphics[width = 1.0\linewidth]{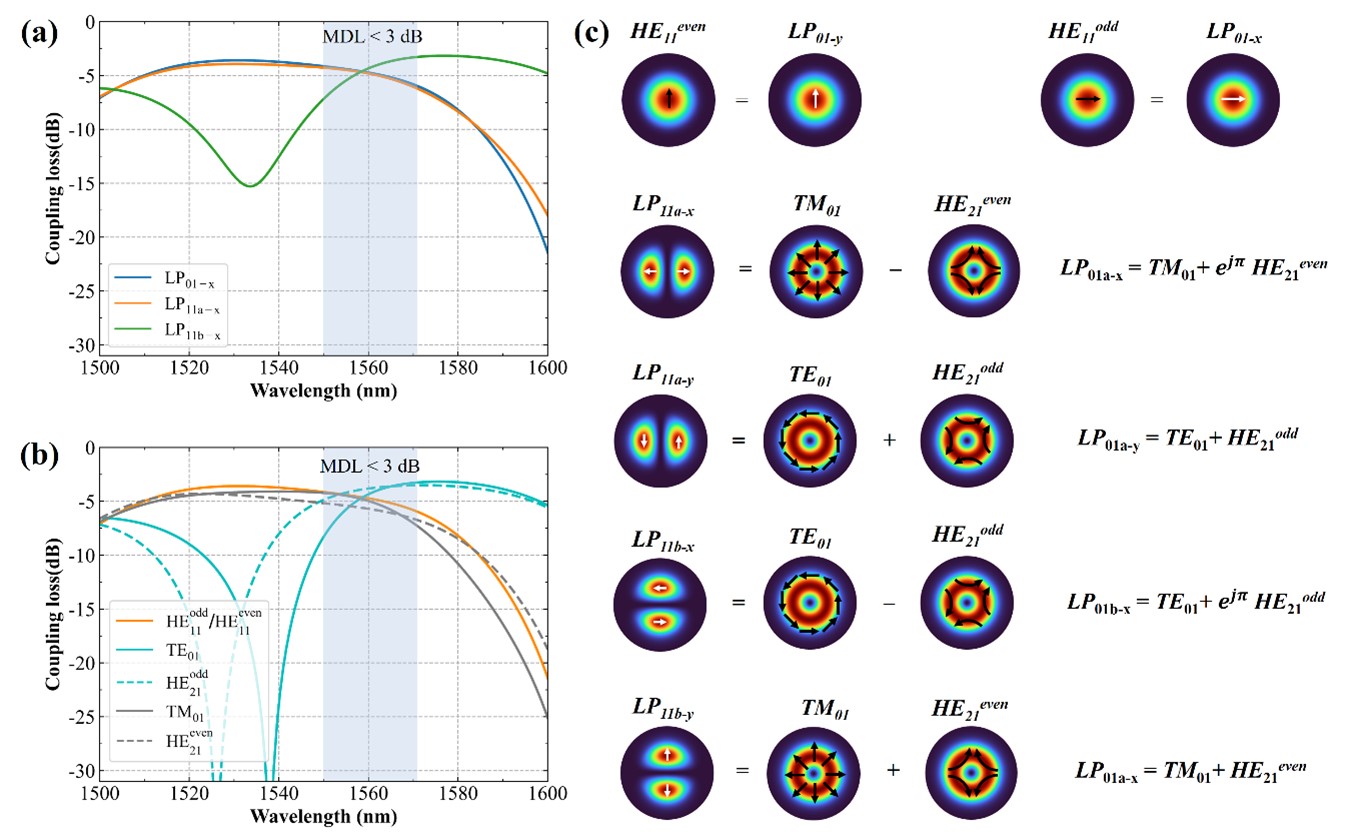}
  } 
  \caption{(a) Chip-to-fiber coupling efficiency spectra by launching various LP modes in a two-mode FMF at the transmitter side. (b) Fiber-to-chip coupling efficiency spectra by launching various eigenmodes in a two-mode FMF at the receiver side. (c) Correspondence between the LP modes and eigenmodes in a two-mode FMF.}
 \label{FigureS2}
\end{figure*} 

\begin{figure*}[!hp]
  \renewcommand{\figurename}{Supplementary Figure}
  \centering{
  \includegraphics[width = 1.0\linewidth]{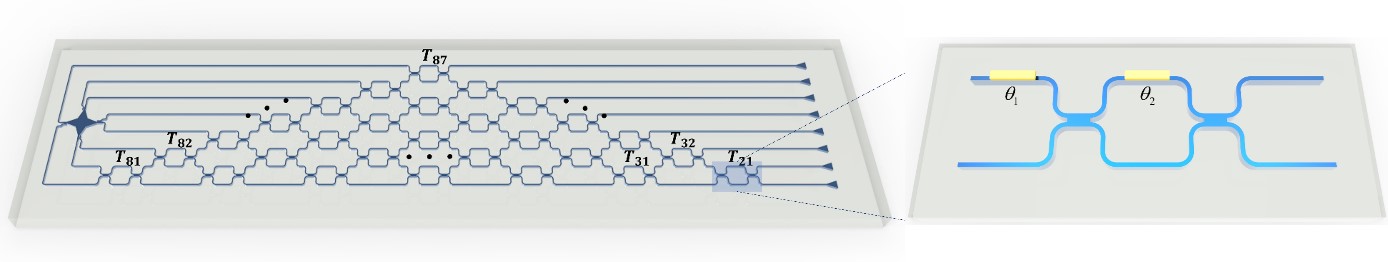}
  } 
  \caption{8$\times$8 unitary optical mesh \textit{U(8)} based on a triangular array of 28 tunable 2$\times$2 beam splitters, which are shown by the inset with two heater-based optical phase shifters.}
 \label{FigureS4}
\end{figure*} 

\begin{figure*}[!hp]
  \renewcommand{\figurename}{Supplementary Figure}
  \centering{
  \includegraphics[width = 1.0\linewidth]{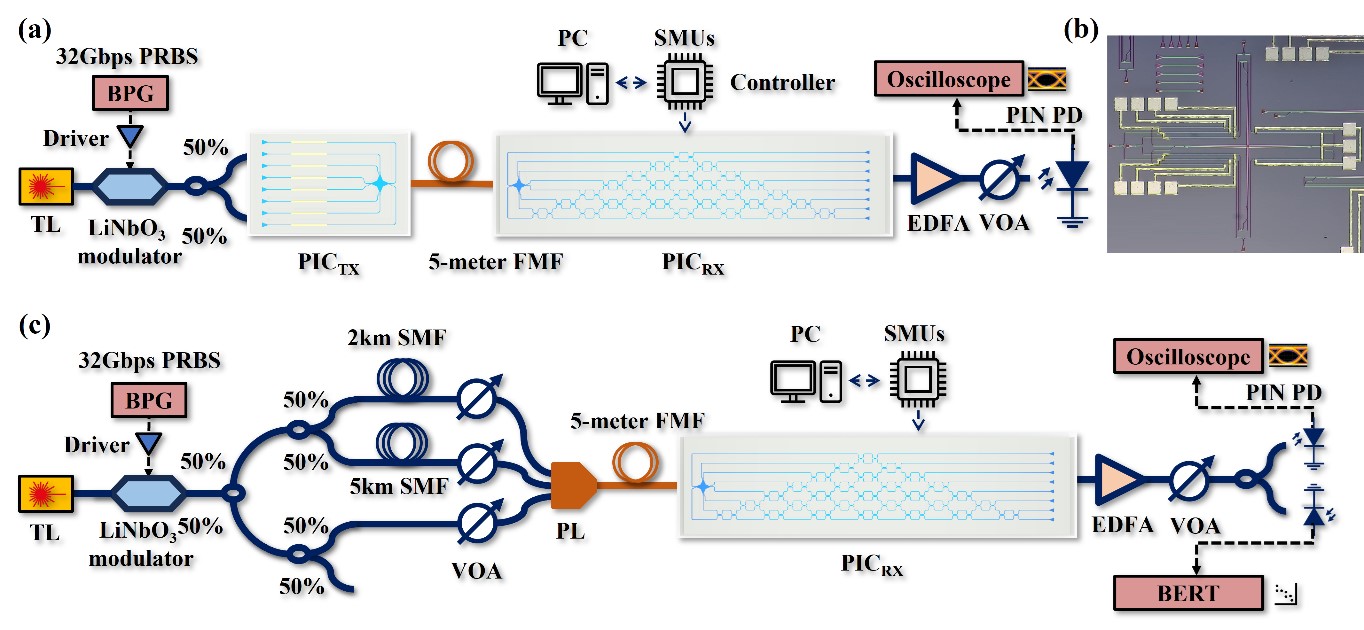}
  } 
  \caption{(a) Experimental setup of inter-chip mode-division multiplexing optical fiber communication system with integrated photonic processor. (b) Microscopic image of the photonic chip at the transmitter side. (c) Experimental setup for evaluation of all-optical mode descrambling with three spatially decoupled fiber modes including LP\textsubscript{01}, LP\textsubscript{11a}, and LP\textsubscript{11b} injected mountainously. TL: tunable laser; BPG: bit pattern generator; SMF: standard single mode fiber; FMF: few-mode fiber; PC: personal computer; SMUs: multichannel source-measurement units; EDFA: erbium-doped fiber amplifier; VOA: variable optical attenuator; PD: photodiode; PL: mode-selective photonic lantern. PRBS: pseudorandom binary sequence; BERT: bit error rate tester; PIC\textsubscript{TX}: photonic integrated circuits at the transmitter side; PIC\textsubscript{RX}: photonic integrated circuits at the receiver side.}
 \label{FigureS5}
\end{figure*} 
\clearpage

\renewcommand{\bibpreamble}{
$^\ast$These authors contributed equally to this work.\\
$^\dagger${Corresponding author: \textcolor{magenta}{hktsang@ee.cuhk.edu.hk}}\\
$^\ddag${Corresponding author: \textcolor{magenta}{yeyutong@hkust-gz.edu.cn}}
}

\bibliographystyle{naturemag}
\bibliography{Reference_supplementary}